\documentclass{PoS}
\usepackage{amstext}
\usepackage{amsmath}
\pdfoutput=1
\title{Comparison of filtering methods in SU(3) lattice gauge theory}

\ShortTitle{Comparison of filtering methods}

\author{F. Bruckmann$^{a}$, \speaker{F. Gruber}$^{a}$, C. B. Lang$^{b}$, M. Limmer$^{b}$, T. Maurer$^{a}$, A. Sch\"afer$^{a}$ and S. Solbrig$^{a}$\\
	\llap{$^a$} Institut f\"ur Theoretische Physik, Universit\"at Regensburg,\\
D-93040 Regensburg, Germany\\
	\llap{$^b$} Institut f\"ur Physik, FB Theoretische Physik, Karl-Franzens-Universit\"at Graz,\\
A-8010 Graz, Austria\\ \newline
        E-mail:	\email{falk.bruckmann@physik.uni-regensburg.de},\\
		\email{florian.gruber@physik.uni-regensburg.de},\\
		\email{christian.lang@uni-graz.at},\\
		\email{markus.limmer@uni-graz.at},\\
		\email{thilo.maurer@physik.uni-regensburg.de},\\
		\email{andreas.schaefer@physik.uni-regensburg.de},\\
		\email{stefan.solbrig@physik.uni-regensburg.de}}

\abstract{
We systematically compare filtering methods used to extract topological excitations from lattice gauge configurations. We show that there is a strong correlation of the topological charge densities obtained by APE and Stout smearing. Furthermore, a first quantitative analysis of quenched and dynamical configurations reveals a crucial difference of their topological structure: the topological charge density is more fragmented, when dynamical quarks are present. This fact also
 implies that smearing has to be handled with great care, not to destroy these characteristic structures.}

\FullConference{8th Conference Quark Confinement and the Hadron Spectrum \\
		 September 1-6 2008\\
		 Mainz, Germany}

\begin{document}

\section{Filtering methods}
Many methods have been developed to extract the IR content from lattice data. Unfortunately, all these methods introduce ambiguities and parameters. Thus, to get a coherent picture of the topological structure of the QCD vacuum, it is necessary to find ways of controlling or even removing these ambiguities.

One of the first attempts to filter out the UV ``noise'' has been APE smearing \cite{APE}, defined as:
\begin{equation}
 U_\mu^{\text{APE}}= P_{SU(N_c)}\left\{(1-\alpha_{APE}) U_\mu^{\text{old}} + \frac{\alpha_{APE}}{6}(\text{staples})\right\},
\end{equation}
where $\alpha_{APE}$ determines the weight of the old link and the sum of the attached staples. The right hand side has to be projected back to the gauge group. Unfortunately, there is no unique mapping. One approach is to take the unitary part of the polar decomposition and normalize this matrix by its determinant. Stout smearing \cite{Morningstar2004a} circumvents this projection by using the exponential map:
\begin{equation}
 U_\mu^{\text{Stout}}= \exp\Big\{ \frac{i}{2} Q_{\mu}(U ,\rho_{\mu\nu})\Big\}\cdot U_\mu^{\text{old}},
\end{equation}
where $Q_{\mu}(U ,\rho_{\mu\nu})$ is a hermitian matrix constructed from all plaquettes containing the old link $U_{\mu}$ and weighted by factors $\rho_{\mu\nu}$. We use the common choice $\rho_{\mu\nu}=\rho_{Stout}$ for isotropic smearing.

A relatively new method is Laplace filtering \cite{Bruckmann2006a}. The filtered links are obtained from a spectral sum of the lowest eigenmodes of the covariant lattice Laplacian\footnote{The original link is reproduced for all eigenmodes, $N=N_c\cdot Vol$, with no projection needed.}:
\begin{equation}
 U_\mu^{\text{Laplace}}(x) = P_{SU(N_c)}\left\{-\sum_{n=1}^{N}\lambda_{n}\Phi_n(x)\otimes\Phi_n^{\dagger}(x+\hat\mu)\right\}.
\end{equation}
This procedure acts as a low-pass filter in the sense of a Fourier decomposition. At this point it should be stressed that Laplace filtering is completely different from smearing, because it is based on rather global objects, namely the eigenmodes, and does not locally modify the gauge links in contrast to smearing.

Taking the filtered links as a starting point, one can reconstruct the topological charge density 
$q(x)=\operatorname{Tr}\big(F_{\mu\nu}(x)\widetilde F_{\mu\nu}(x)\big)/16\pi^2$
from an improved field strength tensor \cite{Bilson-Thompson2003}. 

Also the fermionic definition of the topological charge, via the eigenmodes $\psi$ of a chiral Dirac operator, has been used to explore the IR structure \cite{Diracfilter}. For this so called Dirac filtering one truncates the sum in 
\begin{equation}\label{diracfilter}
q_{Dirac}(x)=\sum_{n=1}^{N} \bigg(\frac {\lambda_n}{2}-1\bigg)\psi^\dagger_n(x)\gamma_5\psi_n(x)
\end{equation}
 and takes only the lowest $N$ modes into account. While the zero-modes determine the total topological charge $Q=\sum q(x)$ due to the index theorem, the non zero-modes modify the local structure of the density, leaving the total charge unaffected.
\section{Comparison of the different methods}
In an earlier study a qualitative and quantitative similarity of the introduced filtering methods for quenched SU(2) gauge configurations has been observed \cite{Bruckmann2007c}. One central element of this comparison is the correlator of two topological charge densities $q_A(x)$ and $q_B(x)$ defined by:
\begin{equation}
\chi_{A B}\equiv\big(1/V\big)\sum_x\;\big(q_A(x)-\overline{q}_A\big)\;\big(q_B(x)-\overline{q}_B\big),
\end{equation}
where the mean values are subtracted for convenience. From this we can construct a quantity that reflects the ``matching'' of two methods:
\begin{equation}
\Xi_{A B}\equiv \frac{ \chi_{A B}^2}{ \chi_{A A}\;\chi_{B B}}
\end{equation}
$\Xi_{A B}$ is obviously equal to one, if $q_A(x)$ is proportional to $q_B(x)$ and deviates the more from one, the more the densities differ.

\begin{figure}[ht]
 \centering
\begin{tabular}{cc}
\begin{minipage}[l]{0.45\textwidth}
 \includegraphics[width=\textwidth]{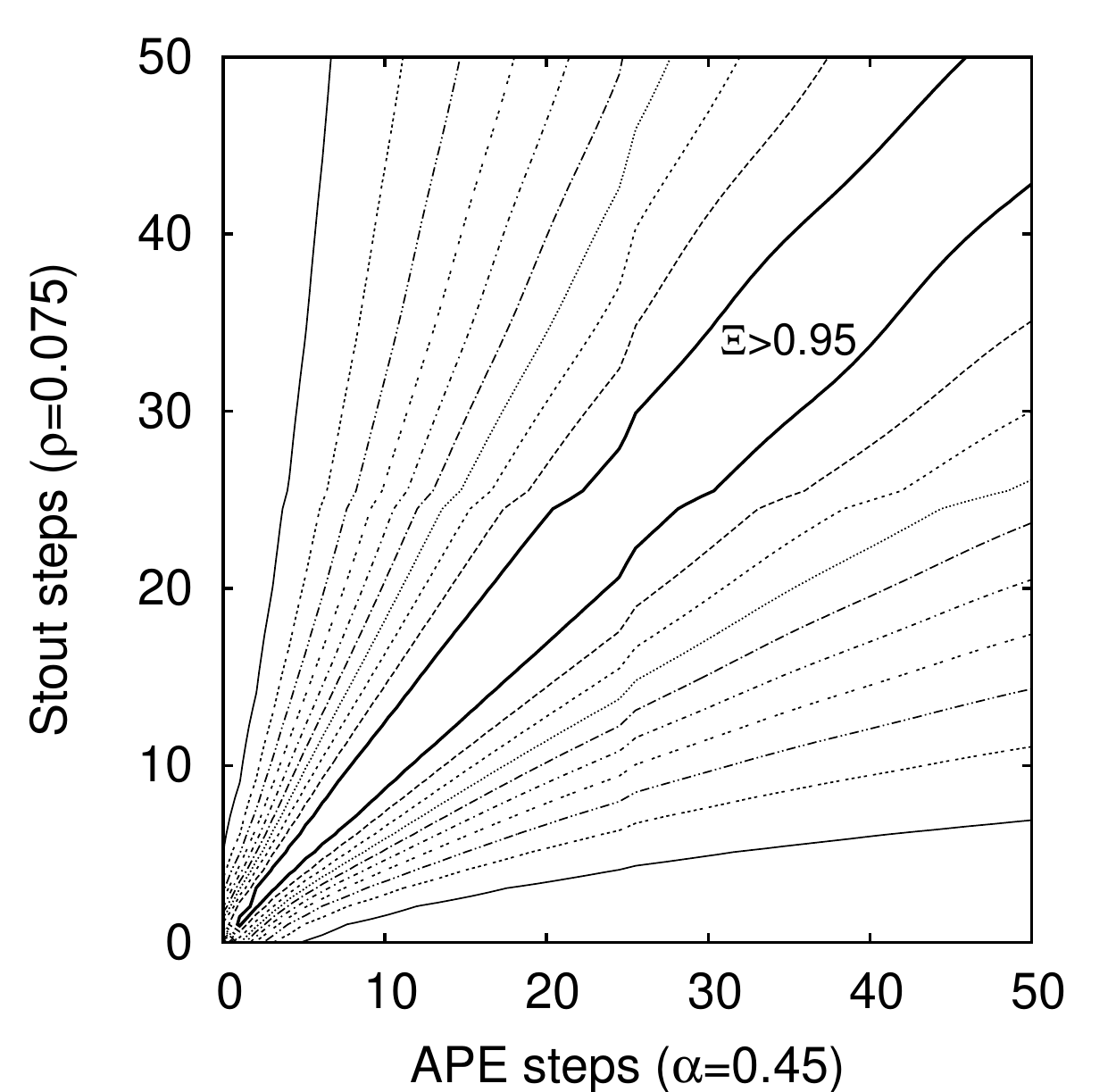}
\end{minipage} &  \begin{minipage}[r]{0.45\textwidth}
\mbox{}\\[3.5em]
 \includegraphics[width=\textwidth]{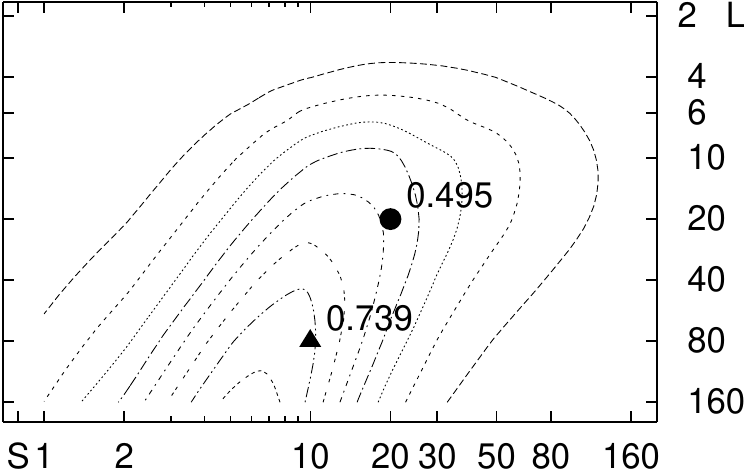}
\end{minipage}
 \end{tabular}
\caption{Level curves of $\Xi = 0.95,0.85,\ldots$ (starting from the diagonal) for APE vs. Stout smearing (left) and $\Xi = 0.8,0.7,\ldots$ (starting from the inside) for APE smearing (S) vs. Laplace modes (L) (right). $\blacktriangle$ and $\bullet$ mark two examples of ``matching'' parameters  for weak and strong filtering respectively (from \cite{Bruckmann2007c}).} \label{fig:Xiplots}
\end{figure}

The main idea is now to relate different filter parameters for those combinations where $\Xi$ is maximal. In fig.\ \ref{fig:Xiplots} the contour lines of $\Xi$ for several methods and parameter ranges are shown. On the right hand plot two exemplary combinations are indicated that correspond to the best matching value for different filtering strengths. 

An interesting observation is that there is an almost one-to-one correspondence for n steps of APE and n steps of Stout smearing when $\alpha_{APE}\approx6\cdot\rho_{Stout}$. As seen in the plot on the lhs.\ of fig.\ \ref{fig:Xiplots}, $\Xi > 0.95$ for a large number of smearing steps. This is consistent with results by Capitani et al.\ \cite{Capitani2006e}, where such a relation has been derived from perturbation theory. While they have focused on global observables with up to 3 smearing steps, our nonperturbative result reflects the local similarity of both methods and their strongly correlated topological charge densities up to 50 steps.
\vspace*{-.8em}
\section{Cluster analysis of the topological charge density}
Another important challenge is to extract observables from lattice data, that could be compared with continuum models of the vacuum. One possibility is to analyze the cluster structure of the topological charge density. Two lattice points belong to the same cluster, if they are nearest neighbors and have the same sign of the topological charge density. Bruckmann et al.\ \cite{Bruckmann2007c} found a power law for the number of clusters as function of the ratio of points with $|q(x)|$ lying above a variable cut-off $q_{cut}$ and the total number of lattice points.
The exponent $\xi$ of this power law is highly characteristic for the topological structure of the QCD vacuum. Different models lead to different predictions, which allows for a very sensitive test. If one has for instance pure noise, the exponent is $1$, as every point forms its own cluster. On the other hand one will have an exponent close to zero for very smooth densities with large structures.
\begin{figure}[!t]
\begin{tabular}{c}
\hspace*{-0.65cm}
\begin{minipage}[l]{\textwidth}
\includegraphics[width=0.54\textwidth]{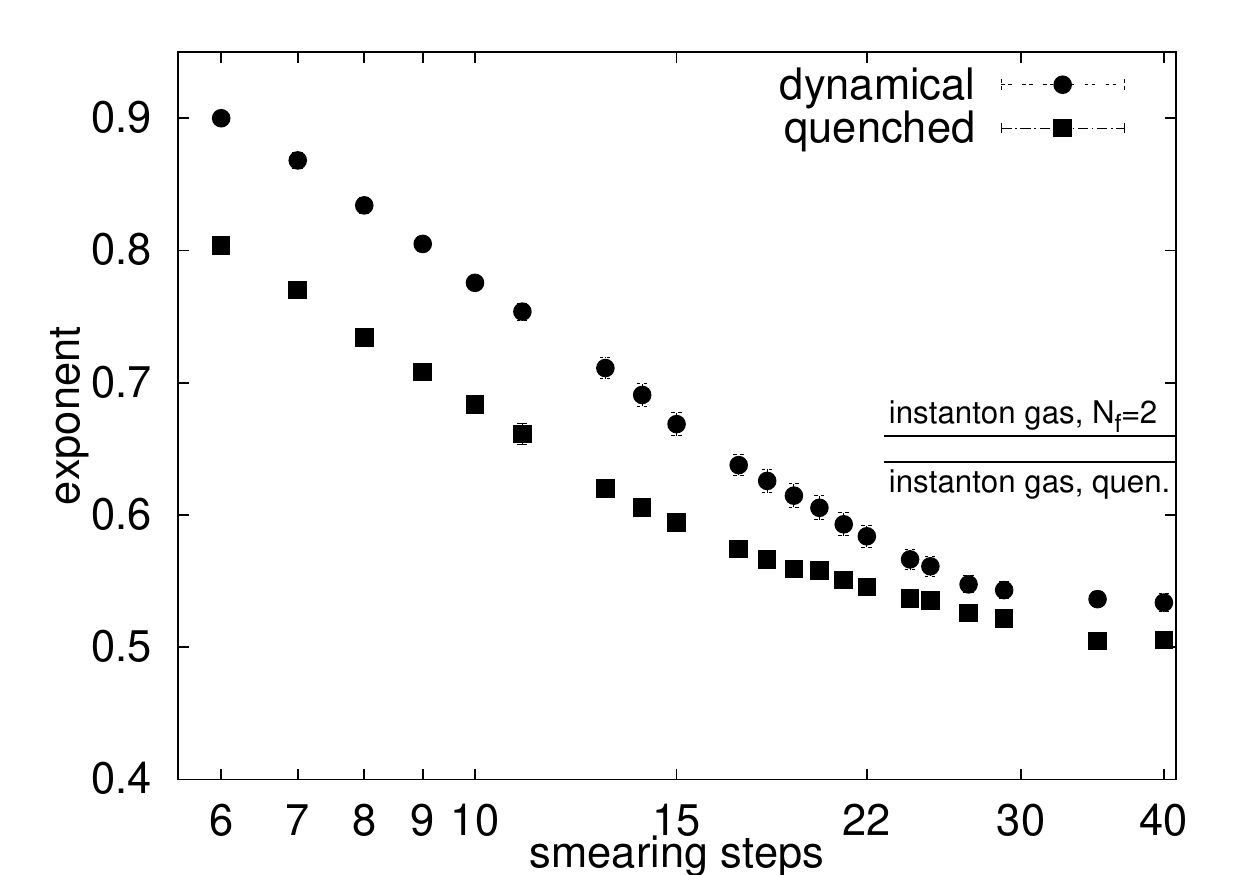}
\hspace*{-1.5cm}
\includegraphics[width=0.57\textwidth]{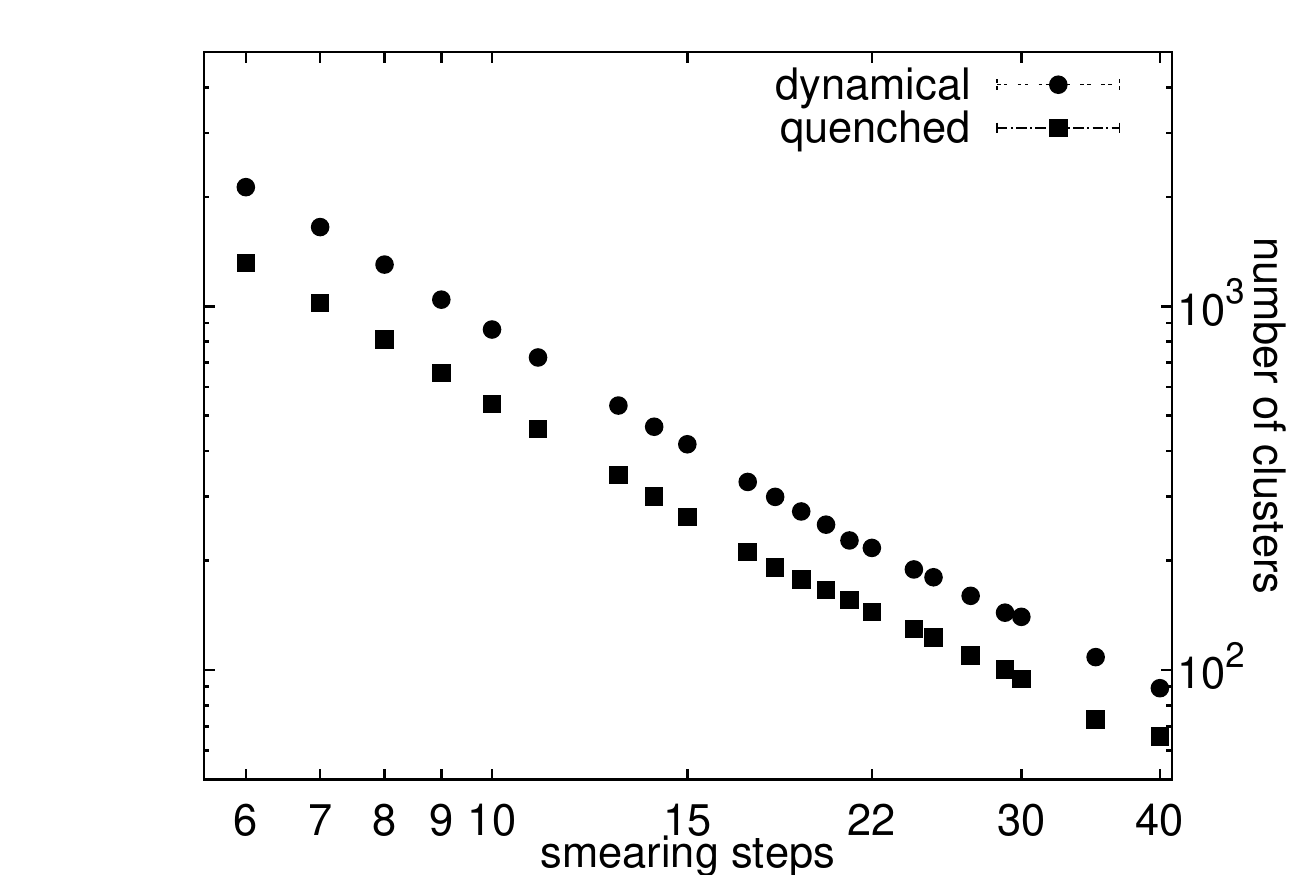}
\end{minipage}
\end{tabular}
\caption{left: Exponent $\xi$ of the analysis for clusters common to APE and Stout smearing. The solid lines show the values predicted from the dilute instanton gas. right: Total number of distinct clusters for a constant fraction $f=0.0755$ of points lying above the cut-off.  Less than 6 steps are not considered, as the definition of the topological charge gets ill-defined. Errors have been calculated using an ensemble average over 10 configurations but are partly too small to see.}\label{fig:expsu3.pdf}
\end{figure}%

To reduce ambiguities we take only those clusters into account, which are common to different filters, whose parameters were matched according to maximal values of $\Xi$. So, if there is an artifact coming from one method, it is unlikely that this artifact will also be seen by the other.

The exponent for clusters common to APE and Stout can be found in fig.\ \ref{fig:expsu3.pdf} (left). We used one quenched and one dynamical $N_f=2$ ensemble with equal lattice spacing (see tab.\ \ref{tab:config}). Obviously the exponents of the dynamical configurations lie above the quenched values.

In order to interpret the cluster exponent, a model of dilute quantized topological objects of general shape 
and with a size distribution $d(\rho)\sim \rho^\beta$ has been considered in \cite{Bruckmann2007c}. It leads to $\xi=(1+4/(\beta+1))^{-1}$ (in 4 dimensions). Following this model, our findings give a larger coefficient $\beta$ in the dynamical case. Hence, smaller topological objects become suppressed.

Moreover, the rhs.\ of fig.\ \ref{fig:expsu3.pdf} shows that for a fixed number of points, lying above the cut-off, much more clusters are found in the dynamical case.
Thus we conclude that when fermion loops are taken into account, the topological structure is more complex and fragmented, in the sense of larger number of distinct objects per volume.
This seems to be in accordance to the findings of the Adelaide group, where small instantons have been seen to be suppressed in the presence of dynamical quarks, while the total number of instantons increased, see fig.\ 6 in \cite{Leinweber}.

The difference of the cluster exponents quenched vs.\ dynamical vanishes for stronger smearing ($\sim$ 30 steps) 
and the exponents settle down to the same plateau.
So we have reasons to believe that too much smearing destroys the impact of dynamical quarks.

On the lhs.\ of fig.\ \ref{fig:expsu3.pdf} we have included for comparison the exponents $\xi=7/11\approx0.64$ and $\xi=23/35\approx0.66$ for the SU(3) instanton gas without resp.\ with dynamical quarks. Taking the dilute instanton gas as a simplified model, it is obvious that the true vacuum should have a higher exponent, as more structures are present. However, the result in fig.\ \ref{fig:expsu3.pdf} (left) shows that this is only the case for very few smearing steps, for slightly
stronger filtering we reach smoother configurations than predicted by the dilute instanton gas. This is another indication of smearing artefacts.%
\begin{table}[htb]
 \begin{center}
\begin{tabular}{lcccc}
\hline\hline
 & lat. size & lat. spacing & $\beta_{LW}$ & $m_0$ \\ \hline
quenched & $16^3\cdot32$ &  0.148 & 7.90 &  --\\ 
dynamical & $16^3\cdot32$ & 0.150 & 4.65 & -0.060\\ \hline\hline 
\end{tabular}
 \end{center}
\caption{Ensembles were generated with the L\"uscher-Weisz gauge action and a chirally improved Dirac operator \cite{CIref}. For the dynamical simulations two flavors of mass degenerate light quarks were used \cite{CIdyn}.}\label{tab:config}
\end{table} 
\section{Conclusion and outlook}
In conclusion, we have found a strong correlation of the topological charge densities obtained from APE and Stout smearing. Furthermore, our first results for dynamical quarks imply that the topological structure is more complex and fragmented in the presence of fermion loops. But there are also indications that smearing has to be used with great caution, especially when dealing with dynamical configurations. The smallness of the cluster exponent is a sign that smearing is more destructive in SU(3) than in SU(2). Preliminary results for clusters common to APE smearing and Laplace filtering do not show such artefacts, as Laplace filtering preserves smaller objects better. This effect is under investigation.


\begin{thebibliography}{99}

\bibitem{APE}
M.~Falcioni et~al., 
 \newblock {\em Nucl.\ Phys.} \textbf{B251} (1985) 624;\
M.~Albanese et~al.,
 \newblock {\em Phys.\ Lett.} \textbf{B192} (1987) 163 

\bibitem{Morningstar2004a}
C. Morningstar and M.~J. Peardon,
 \newblock {\em Phys. Rev.}, \textbf{D69} (2004) 054501,
 \newblock [{\tt{\href{http://arxiv.org/abs/hep-lat/0311018}{hep-lat/0311018}}}]

\bibitem{Bruckmann2006a}
F. Bruckmann and E.~M. Ilgenfritz.
 \newblock {\em Phys. Rev.} \textbf{D72} (2005) 114502,
 \newblock [{\tt{\href{http://arxiv.org/abs/hep-lat/0509020}{hep-lat/0509020}}}]

\bibitem{Bilson-Thompson2003}
S.~O. Bilson-Thompson et~al.,
 \newblock {\em Annals.\ Phys.} 304 (2003) 1-21,
 \newblock [{\tt \href{http://arxiv.org/abs/hep-lat/0203008v1}{hep-lat/0203008v1}}]

\bibitem{Diracfilter}
P.~Hasenfratz, V.~Laliena and F.~Niedermayer,
 \newblock {\em Phys.\ Lett. }  {\bf B427} (1998) 125,
 \newblock [{\tt{\href{http://arxiv.org/abs/hep-lat/9801021}{hep-lat/9801021}}}];\
I.~Horvath et~al.,
\newblock {\em Phys.~Rev.} \textbf{D67} (2003) 011501,
\newblock [{\tt{\href{http://arxiv.org/abs/hep-lat/0203027}{hep-lat/0203027}}}];\
I.~Horvath et~al.,
\newblock {\em Phys.~Rev.} \textbf{D68} (2003) 114505,
\newblock [{\tt{\href{http://arxiv.org/abs/hep-lat/0302009}{hep-lat/0302009}}}]

\bibitem{Bruckmann2007c}
 F. Bruckmann et~al.,
 \newblock \emph{Eur. Phys. J}., A33:333--338 (2007),
 \newblock [{\tt{\href{http://arxiv.org/abs/hep-lat/0612024v3}{hep-lat/0612024}}}]

\bibitem{Capitani2006e}
 S.~Capitani, S.~D\"urr, and C. Hoelbling,
 \newblock {\em JHEP}, 11:028 (2006),
 \newblock [{\tt \href{http://arxiv.org/abs/hep-lat/0607006}{hep-lat/0607006}}]

\bibitem{Leinweber}
D.~Leinweber and P.~J. Moran,
 \newblock {\em Phys.~Rev.} \textbf{D78} (2008) 054506,
 \newblock {[\tt \href{http://arxiv.org/abs/0801.2016v1}{arxiv:0801.2016}]}

\bibitem{CIref}
C.~Gattringer,
 \newblock {\em Phys.\ Rev.} {\bf D63} (2001) 114501,
 \newblock [{\tt \href{http://arxiv.org/abs/hep-lat/0003005}{hep-lat/0003005}}];
  C.~Gattringer, I.~Hip and C.~B.~Lang,
 \newblock { \em Nucl.\ Phys.} {\bf D597} (2001) 451,
 \newblock [{\tt \href{http://arxiv.org/abs/hep-lat/0007042}{hep-lat/0007042}}]
\bibitem{CIdyn}
C.~B.~Lang et~al.,
 \newblock {\pos{PoS(LATTICE 2007)114} (2007)},
C.~B.~Lang et~al.,\newblock {(2008), [{\tt \href{http://arxiv.org/abs/0812.1681}{hep-lat/0812.1681}}]}

\end{thebibliography}
\end{document}